\documentclass[pra,twocolumn,showpacs]{revtex4}
\usepackage{amsmath}
\usepackage{amssymb}
\usepackage{epsfig}

\begin{document}

\title{\bf On the nonlinear Schr\"odinger equation for waves on a nonuniform current}
\author{V.~P. Ruban}
\email{ruban@itp.ac.ru}
\affiliation{L.D. Landau Institute for Theoretical Physics RAS, Moscow, Russia} 

\date{\today}

\begin{abstract}
A nonlinear Schr\"odinger equation with variable coefficients for surface waves on a large-scale steady 
nonuniform current has been derived without the assumption of a relative smallness of the velocity of the current.
This equation can describe with good accuracy the loss of modulation stability of a wave coming to a counter
current, leading to the formation of so called rogue waves. Some theoretical estimates are compared to the
numerical simulation with the exact equations for a two-dimensional potential motion of an ideal fluid with
a free boundary over a nonuniform bottom at a nonzero average horizontal velocity.
\end{abstract}

\maketitle

The nonlinear Schr\"odinger equation (NLSE) plays an important role in the theory of waves of various
natures. In application to waves on the free surface of a deep fluid in the gravitational field, it was first
derived by V. E. Zakharov [1]. In particular, this equation describes the modulation instability of a plane
wave owing to focusing nonlinearity [1, 2]. The nonlinear Schr\"odinger equation has recently attracted
increased interest in view of active investigations of so-called anomalous waves (in oceanography, they are
called rogue or freak waves; see [3-6] and references therein). The direct numerical simulation with the
exact equations of motion of an ideal fluid with the free surface indicates that modulation instability is a
crucial factor for the appearance of anomalous waves [7-9]. Rogue waves appear at the nonlinear stage of
the development of this instability, when the mechanical energy is concentrated at a scale of one or two
wavelengths.

It is known that a wave with carrier wavenumber $k_0$, modulation wavenumber $\Delta k$, and amplitude $A_0$ is
modulation unstable under the condition $\varepsilon_0 N>1/(2\sqrt 2)$, 
where $\varepsilon_0=k_0 A_0$ is the wave steepness and $N=k_0/\Delta k$ is the number of wavelengths under one
spatial modulation period (in some works, $N$ is defined as the number of periods of waves under one time
modulation period at a fixed spatial point; the number $N$ thus defined differs by a factor of 2). In the presence
of only one perturbation mode, the development of instability is well described by the so called Akhmediev breather 
that is an exact solution of the nonlinear Schr\"odinger equation [10-13], which gives the following estimate for 
the maximum amplitude of an anomalous wave:
\begin{equation}
\label{A_max}
A_{\rm max}/A_0=1+2\sqrt{1-(2\sqrt{2}\varepsilon_0 N )^{-2}}.
\end{equation}

If several perturbation modes with different $(\Delta k)_i$ values exist at the initial time, the analytical 
solution is also possible, but it is very complicated (see [14] and references therein).

For random wave fields with average steepness $\varepsilon$, whose spectrum is concentrated in the wavenumber
range with a width of $\Delta k$ near $k_0$ , the so-called Benjamin-Feir index $I\propto \varepsilon k_0/\Delta k$ 
is introduced; it determines the capability of this wave field to generate
anomalous waves owing to nonlinear self-focusing. At small $I$ values, the probability of the appearance of
anomalous waves is small, whereas at $I\sim 1$, there are sufficiently long and/or high groups of waves in which
modulation instability is developed and rogue waves appear.

The product $\varepsilon N$ for usual marine conditions is small. For example, under storm conditions, when the
wave steepness is large, long wave groups are absent and, on the contrary, ocean swell far from a storm at
large $N$ values has a small wave steepness. For this reason, rogue waves rarely appear in oceans. 
The appearance of an anomalous wave usually requires the preliminary action of some additional factors that would
significantly increase the Benjamin-Feir index. One of the most efficient mechanisms for an increase in the
wave steepness is its interaction with a nonuniform current (see, e.g., [15, 16]). For simplicity, 
two-dimensional flows will be considered in this work (anomalous waves in three-dimensional space have
additional specificity; see, e.g., [17-22]). This mechanism is linear and due to the fact that the wavenumber $k$
of each monochromatic component with frequency $\omega$ in the presence of a large-scale nonuniform current
$U(x)$ varies slowly along the horizontal coordinate $x$
according to the local dispersion relation $\omega = Uk+\sqrt{gk} =const$ (let  $k>0$ for definiteness; i.e., the wave
propagates in the positive direction of the $x$ axis). This dispersion relation provides the formula
\begin{equation}
\label{k_omega_U}
k(\omega,U)=[g+2\omega U-\sqrt{g^2+4g\omega U}]/(2U^2).
\end{equation}
The amplitude $A_\omega$  of this wave component satisfies the
conservation law of the wave action [23]:
\begin{equation}
\label{wave_action_conservation}
\Big[U+(1/2)\sqrt{g/k}\Big]g|A_{\omega}|^2/\sqrt{gk}\approx\mbox{const}. 
\end{equation}
It follows from Eqs. (2) and (3) that $A_\omega(x)\propto M(x,\omega)$, where
\begin{equation}
\label{A_propto_M}
M \approx (k/g)^{1/2}[1+4\omega U/g]^{-1/4}. 
\end{equation}
At  $U<0$, i.e., on the counter current, the steepness $\tilde\varepsilon=kA_\omega$ increases strongly, 
because both the wavenumber and amplitude increase. A change in the local $N$
value along the $x$ axis is easily estimated from the condition of the conservation of the frequency of each
wave component: $N=k/\Delta k\approx k/(k_\omega \Delta\Omega)$, where $k_\omega$ is
the partial derivative of the function $k(\omega,U)$ with
respect to $\omega$, and $\Delta\Omega$ is the constant modulation frequency. 
Substitution and simplification give the formula
\begin{equation}
\label{I}
I_\upsilon/I_0=16[1+\sqrt{1+\upsilon}]^{-4}(1+\upsilon)^{\frac{1}{4}},
\end{equation}
where $\upsilon= 4\omega U/g$. If the modulated wave that is in a
modulation stable state in a weak current $\upsilon_1$ comes to
a sufficiently fast counter current $\upsilon_2$, it can prove to be
in a modulationally unstable state. In this case, instead
of Eq. (1), we have the estimate
\begin{equation}
\label{A_max_I}
A_{{\rm max},2}/\langle A_2\rangle\approx
1+2\sqrt{1-[2\sqrt{2}(\varepsilon N)_1 I_2/I_1]^{-2}},
\end{equation}
where angular brackets stand for the average value.

For a more detailed analysis of the dynamics of
waves on a nonuniform current, it is necessary to integrate effects caused by spatial nonuniformity, 
as well as the dispersion and nonlinearity of waves, in one
model. This integration should provide a certain modified nonlinear Schr\"odinger equation; the derivation
of this equation is the main aim of this work. It is worth
noting that attempts to derive the required equation
are known (see [24-27]). However, the proposed variants of the nonlinear Schr\"odinger equation modified
with the inclusion of a nonuniform current have disadvantages or restrictions. In particular, the equations
from [26, 27] in the linear limit give results that
strongly differ (even in the first order in $\upsilon$) from Eq.
(4); i.e., they strongly contradict the principle of the
conservation of the wave action. The reason for the
inconsistency is a methodical error common for [26,
27], where both the amplitude of the wave and (presumably small) variations of the velocity $\tilde U$
expressed in terms of one small parameter $\varepsilon$. However,
these two quantities are independent of each other in
their physical meanings. Under that assumption, the
correct passage to the limit of the small wave amplitude at a given current is impossible. The incorrect
equation from [27] was recently used as a basis model
in [28]. For this reason, the quantitative results
obtained in [28] are erroneous (for example, cf. Eq. (6)
in this work and Eq. (8) in [28]).

After a certain change of the dependent variable,
the equations derived in [24, 25] by different methods
(without the assumption of the smallness of the current velocity) are close to the nonlinear Schr\"odinger
equation that will be derived below. A difference is
observed only in the second-order dispersion terms,
which ensure the variational structure of the equation
obtained in this work and a somewhat wider region of its applicability.

{\bf Derivation of the NLSE}.
Consideration begins with the linear part of the
problem. The effect of nonlinearity on a quasi-monochromatic wave on deep water is mainly reduced to the
nonlinear frequency shift and, consequently, it can be easily included in the equation. The derivation of the
linear part of the equation is nontrivial.

It is convenient to use the Hamiltonian formulation of the dynamics of the free surface of an ideal fluid [1].

In the presence of a (potential) two-dimensional flow, the system has a steady state in which the profile
of the free boundary $y=\eta_0(x)$  satisfies the time-independent Bernoulli equation
$U^2(x)/2+g\eta_0(x)\approx const=\langle U^2(x)/2\rangle$. In this case, the vertical component of the steady 
velocity field on the surface $y=\eta_0(x)$ can be neglected. This velocity field is directed
along a tangent and the corresponding slope of the
tangent in the cases under study is negligibly small, i.e., $|\eta'_0(x)|\ll 1$.

Time-dependent perturbations of the system, i.e., surface waves, will be described in terms of the functions 
$\eta$ and $\psi$, where $\eta(x,t)$ is the vertical deviation
from the steady profile and $\psi(x,t)$ is the surface value
of the velocity field potential created by a wave. As is
known, $\eta$ and $\psi$ constitute a pair of canonically conjugate variables and the Hamiltonian of the system in
this case is the difference between its total energy and
the energy of the steady state.

It is assumed that the $U(x)$ dependence is due to the
inhomogeneity of the bottom profile $h(x)$, which is
sufficiently deep so that the deep water regime for surface waves is ensured, i.e., $\exp(-2kh)\ll 1$. 
The characteristic length $\Lambda$ at which the function $U(x)$ varies
noticeably is assumed to be much larger than the
wavelength, i.e., $\Lambda k\gg 1$ and, hence, $\Lambda \gtrsim h$. Under
these conditions, $U(x)h(x)\approx const$. Such a situation
can occur, e.g., near river mouths or in ebb currents.

Under the accepted assumptions, the quadratic
part of the Hamiltonian of surface waves can be represented in the form
\begin{equation}
{\cal H}^{(2)}=\int\Big[U\eta\psi_x +\frac{g\eta^2}{2}
+\frac{\psi\hat K \psi}{2} +{\cal O}(\Lambda^{-2})\Big]dx,
\end{equation}
where $\hat K$ is the self-adjoint linear operator that multi
plies the Fourier transform $\psi_k$ by $|k|$. Corrections of
the order of $\Lambda^{-1}$ are taken into account, whereas
higher order terms in this parameter are neglected.
The corresponding linear equations of motion have the form
\begin{equation}
\label{eta_t_psi_t}
\eta_t = -(U\eta)_x +\hat K \psi,\qquad
-\psi_t= U \psi_x +g\eta.
\end{equation}

These two real equations can be transformed to one
complex equation for the envelope of a quasi-monochromatic wave. First, strictly monochromatic 
solutions of system (8) will be considered. The substitution
\begin{equation}
\eta=\mbox{Re}[Q(x,\omega)e^{-i\omega t}],\quad
\psi=\mbox{Re}[P(x,\omega)e^{-i\omega t}],  
\end{equation}
and exclusion of $Q$ provide the equation for $P$:
\begin{equation}
\label{P_omega}
\omega^2 P +i\omega(\partial_x U+U\partial_x)P
-\partial_x U^2 \partial_x P=g{\hat K}P.
\end{equation}
Since $U(x)$ varies slightly at a wavelength, the approximate solution of Eq. (10) can be sought in the form
\begin{equation}
P(x,\omega)\approx\Psi(x,\omega)\exp\Big[i\!\int^x\!\! k(\omega,U)dx\Big],
\end{equation}
where $\Psi(x,\omega)$ is a slow function of the coordinate. It
is very important that $k(\omega,U)>0$. For this reason, ${\hat K}P\approx-i\partial_x P$, 
because the Fourier spectrum of the
function $P$  is almost completely concentrated on positive wavenumbers. 
The substitution into Eq. (10) provides the equation
\begin{equation}
\omega (U_x \Psi + 2 U \Psi_x)-2kU(U_x\Psi+U\Psi_x)
-U^2k_x\Psi+g\Psi_x\approx 0.
\end{equation}
Multiplying it by $\Psi$ and integrating with respect to  $x$,
we obtain  $[\omega U-kU^2 +g/2]\Psi^2\approx\mbox{const}$ (in agreement
with the conservation of the wave action) or
\begin{equation}
\Psi\approx -iC [1+4\omega U/g]^{-1/4},
\end{equation}
with an arbitrary complex constant $C$. Since $gQ=i\omega P-UP_x$, we have
\begin{equation}
Q(x,\omega) = CM(x,\omega)\exp\left[i\int^x k(\omega,U)dx\right],
\end{equation}
in complete agreement with Eqs. (2)-(4).

Now, we consider a linear superposition of monochromatic solutions in a narrow frequency range near $\omega$:
\begin{eqnarray}
\label{envelope}
\eta&=&\mbox{Re}\!\int\! d\xi {\tilde C}(\xi)M(x,\omega+\xi)\,
e^{\{-i(\omega+\xi)t+i\int^x k(\omega+\xi,U)dx\}}\nonumber\\
&\approx&\mbox{Re}\big[ \Theta(x,t)M(x,\omega)\,
e^{\{-i\omega t+i\int^x k(\omega,U)dx\}}\big],
\end{eqnarray}
where
\begin{equation}
\label{Theta_def}
\Theta(x,t)=\int d\xi {\tilde C}(\xi)\,
e^{\{-i\xi t+i\int^x[k(\omega+\xi,U)-k(\omega,U)]dx\}}.
\end{equation}
is a complex function. Formula (15) means that the
wave envelope (in the generalized meaning) is $\tilde A(x,t)\approx \Theta(x,t)M(x,\omega)$, 
although the filling exponential is no longer characterized by a constant wavenumber 
because of the spatial nonuniformity of the system.

Since the frequency spectrum ${\tilde C}(\xi)$ is concentrated in a narrow range $\Delta\Omega\ll\omega$
at small $\xi$ values, the difference $[k(\omega+\xi,U)-k(\omega,U)]$  can be expanded in powers
of $\xi$ and the following partial differential equation for $\Theta(x,t)$  in the linear regime 
is thus obtained from Eq. (16):
\begin{equation}
\label{Theta_linear}
-i\Theta_x=i k_\omega \Theta_t -(1/2)k_{\omega\omega}\Theta_{tt}+\cdots.
\end{equation}
Here,
\begin{eqnarray}
k_\omega&=&(4\omega/g)[1+\upsilon+(1+\upsilon)^{1/2}]^{-1},\\
k_{\omega\omega}&=&(2/g)(1+\upsilon)^{-3/2} 
\end{eqnarray}
are the partial derivatives of the function $k(\omega,U)$.
Since dispersion corrections of higher orders are
neglected in Eq. (17), the proposed theory can be valid
only under the condition  $\upsilon \gtrsim -0.8$ (wave blocking
corresponds to $\upsilon_*=-1$).

In order to complete the derivation of the nonlinear Schr\"odinger equation with variable coefficients for
a weakly nonlinear quasi-monochromatic wave on a nonuniform current, the nonlinear frequency shift
should be taken into account in the standard way. For
a fixed $k$ value, it is $\delta\omega\approx\sqrt{gk}k^2|A|^2/2$. For a fixed frequency $\omega$, 
this corresponds to the nonlinear shift of
the wavenumber $\delta k\approx -k_\omega \sqrt{gk}k^2|A|^2/2$. With the use
of the relation $|A|\approx|\Theta| M$, the equation is obtained in the final form
\begin{equation}
\label{NLSE_Theta_x}
i\Theta_x+i k_\omega \Theta_t -\frac{1}{2}k_{\omega\omega}\Theta_{tt}
-\frac{k_\omega k^3 \sqrt{k}}{2\sqrt{g+4\omega U}}|\Theta|^2 \Theta\approx 0,
\end{equation}
which is the main result of this work.

The evolution variable in Eq. (20) is the $x$ coordinate. Since the terms of this equation have different
orders, it can be rewritten with the same accuracy in
another form, where the evolution variable is the time $t$, which is more conventional for Hamiltonian 
systems. Indeed, since $\Theta_x+ k_\omega \Theta_t=0$ in the leading order,
the second-order dispersion term can be represented
in terms of the derivatives with respect to the coordinate taking into account that 
$\partial_t\approx-k_\omega^{-1}\partial_x$. The resulting equation has the form
\begin{equation}
\label{NLSE_Theta_t}
i\Big[k_\omega\frac{\partial}{\partial t}
+\frac{\partial}{\partial x}\Big]\Theta -
\frac{1}{2}\frac{\partial}{\partial x}
\frac{k_{\omega\omega}}{k_\omega^2}\frac{\partial}{\partial x}\Theta
-\frac{k_\omega k^3 \sqrt{k}}{2\sqrt{g+4\omega U}}|\Theta|^2 \Theta\approx 0,
\end{equation}
which has the variational structure with the Lagrangian
\begin{eqnarray}
{\cal L}&=&\int\Big\{i\Theta^* \Big[k_\omega\frac{\partial}{\partial t}
+\frac{\partial}{\partial x}\Big]\Theta -
\frac{1}{2}\Theta^* \frac{\partial}{\partial x}
\frac{k_{\omega\omega}}{k_\omega^2}\frac{\partial}{\partial x}\Theta
\nonumber\\
&&\quad\quad
-\frac{k_\omega k^3 \sqrt{k}}{4\sqrt{g+4\omega U}}(\Theta \Theta^*)^2 \Big\}dx,
\label{Lagrangian_NLSE_Theta_t}
\end{eqnarray}
where $\Theta^*$  is the complex conjugate function.

It is noteworthy that a change of the dependent
variable $\Theta(x,t)=A(x,t)/M$ in Eq. (21) with allowance
for some properties of the function $k(\omega,U)$ makes it
possible to obtain an equation that almost coincides
with Eq. (3.12) from [25]. A difference is observed
only in some second-order dispersion terms, the
inclusion of which is beyond the method used in [25].

Equation (20) (and (21)) can be supplemented by terms responsible for the linear damping of the wave
and for its slow variations along the transverse horizontal coordinate (assuming, as before, that the 
current $U(x)$ is one-dimensional). The corresponding formulas are not presented in this work.

\begin{figure}
\begin{center}
   \epsfig{file=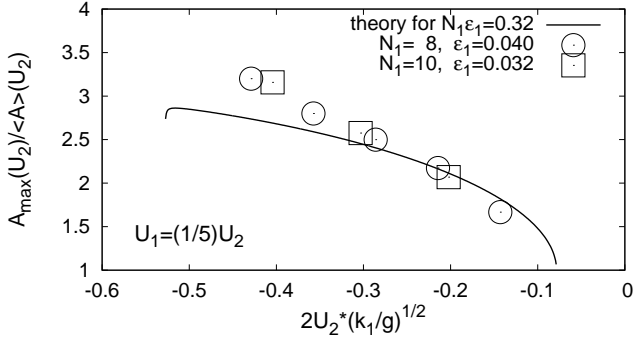,width=88mm}\\
\end{center}
\caption{Theoretical estimate (6) in comparison with numerical results.} 
\label{Ampl} 
\end{figure}

\begin{figure}
\begin{center}
   \epsfig{file=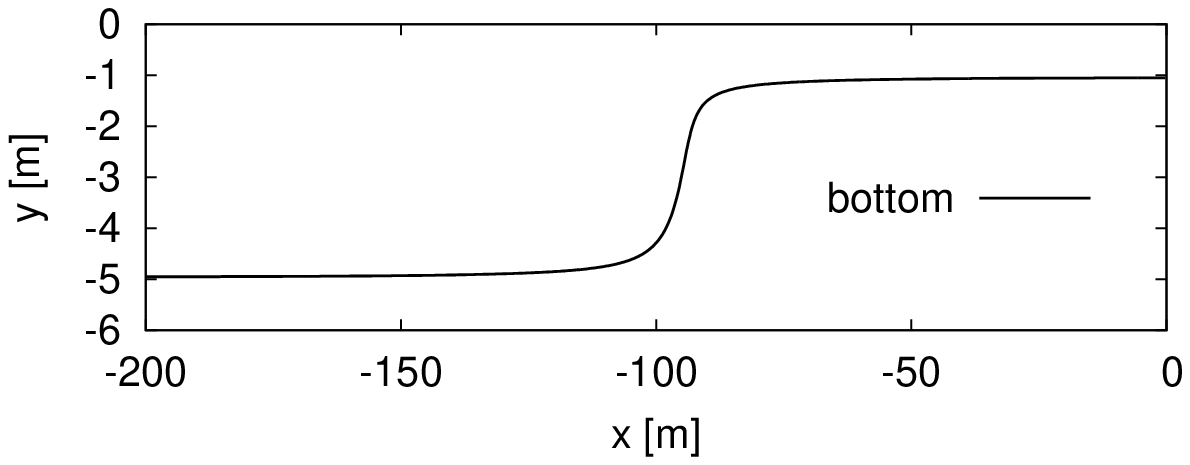,width=88mm}
   \epsfig{file=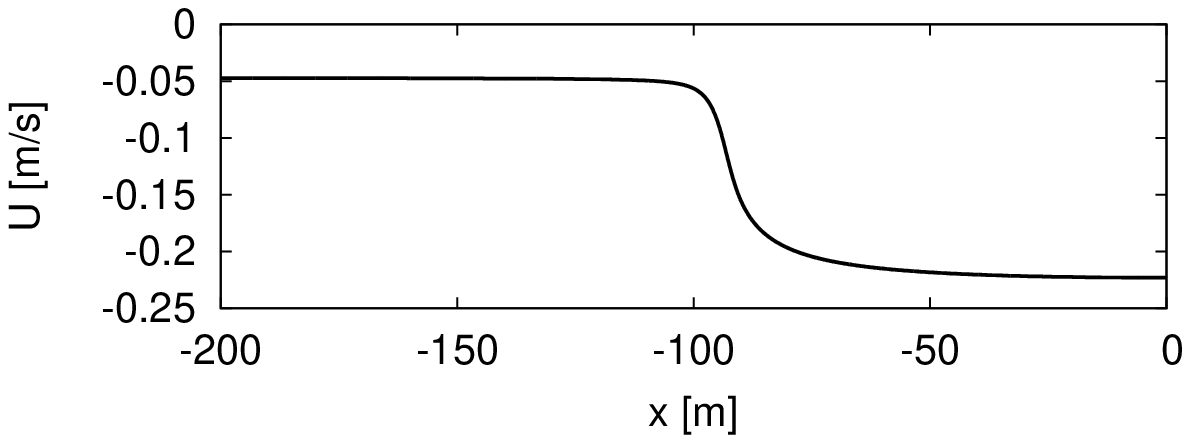,width=88mm}\\
\end{center}
\caption{
(a) Bottom profile given by Eq. (30). (b) Function
$U(x)$ at $s=-0.005$. Only half of the spatial period is shown.} 
\label{U} 
\end{figure}
\begin{figure}
\begin{center}
   \epsfig{file=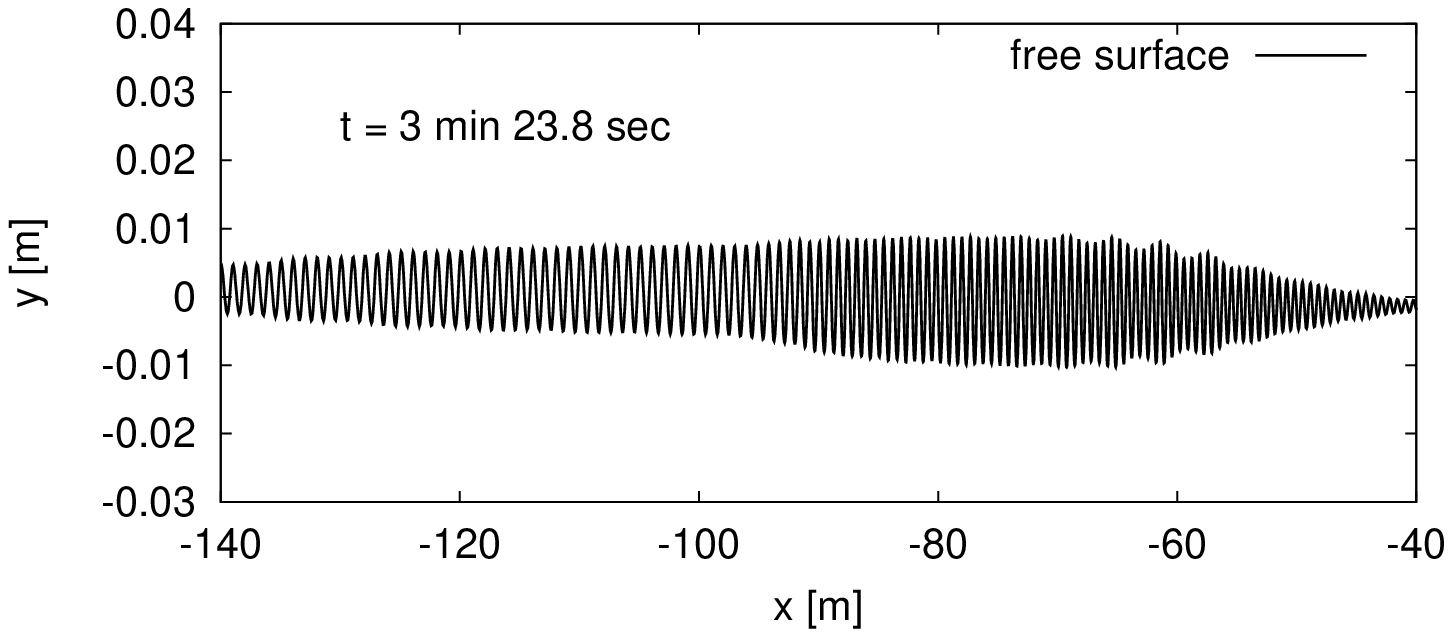,width=88mm}\\
   \epsfig{file=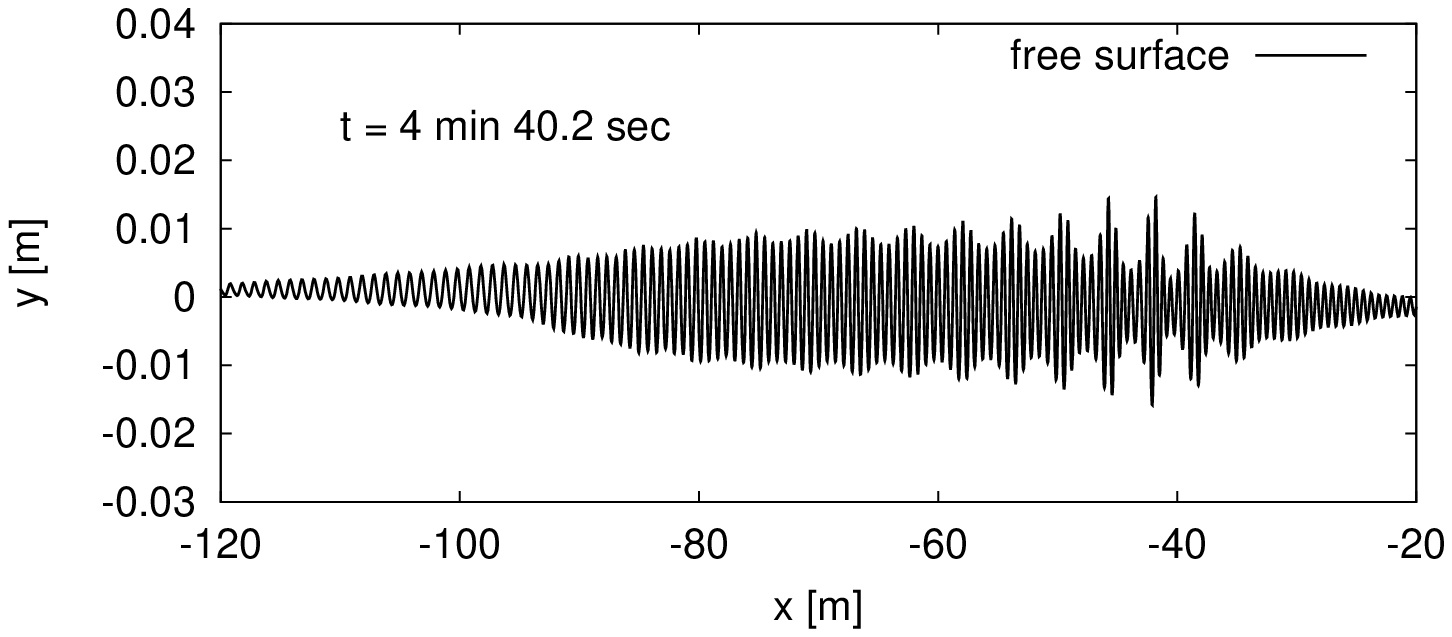,width=88mm}\\
   \epsfig{file=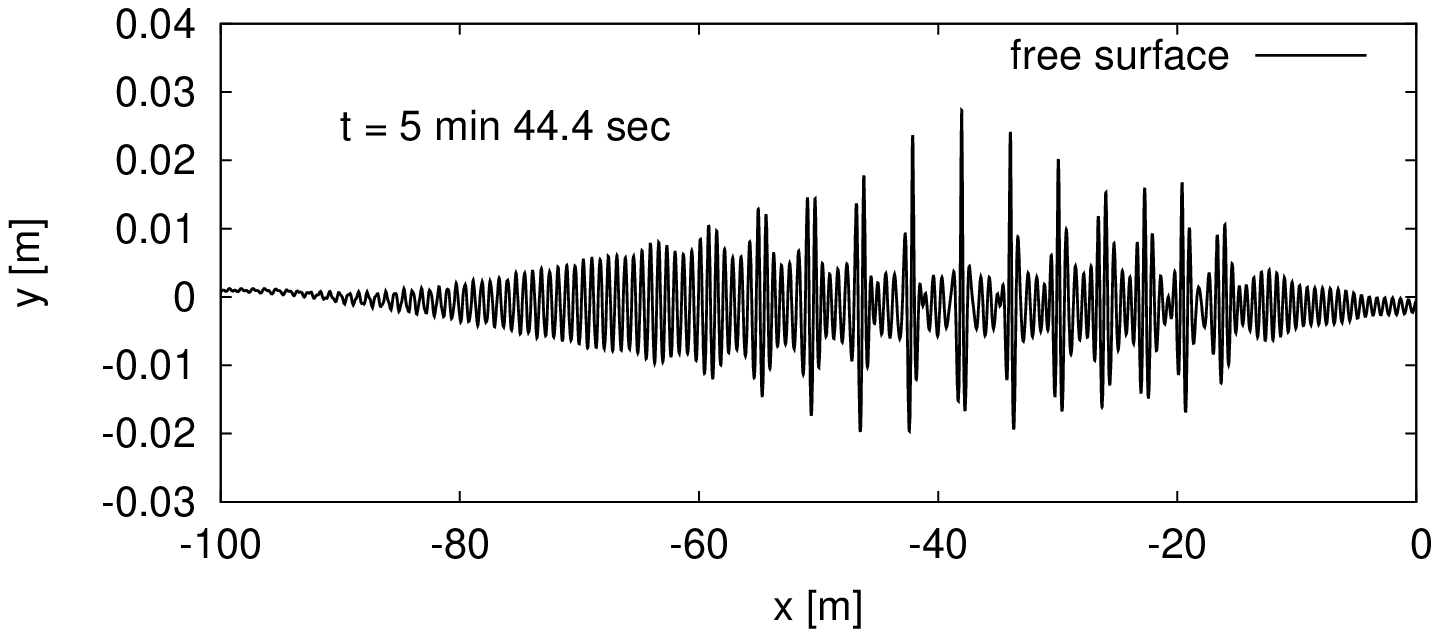,width=88mm}\\
\end{center}
\caption{Example of the development of modulation instability on the counter current.} 
\label{MI_development} 
\end{figure}
\begin{figure}
\begin{center}
   \epsfig{file=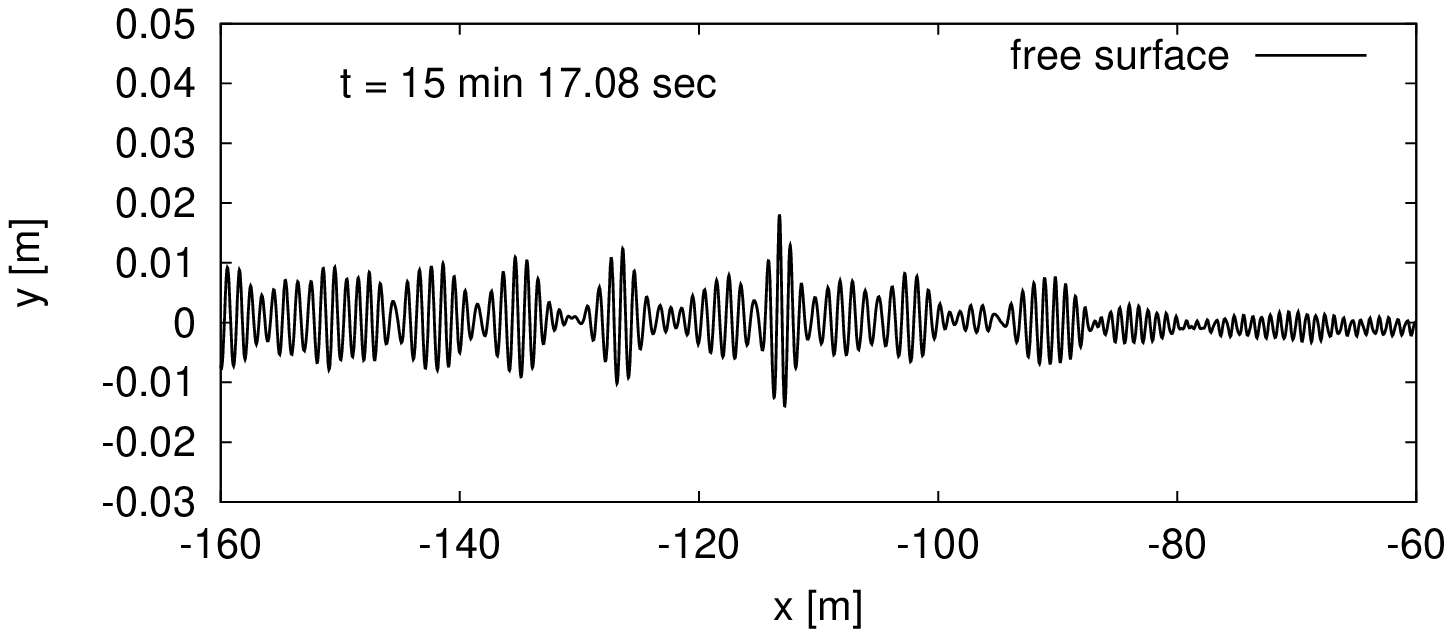,width=88mm}\\
   \epsfig{file=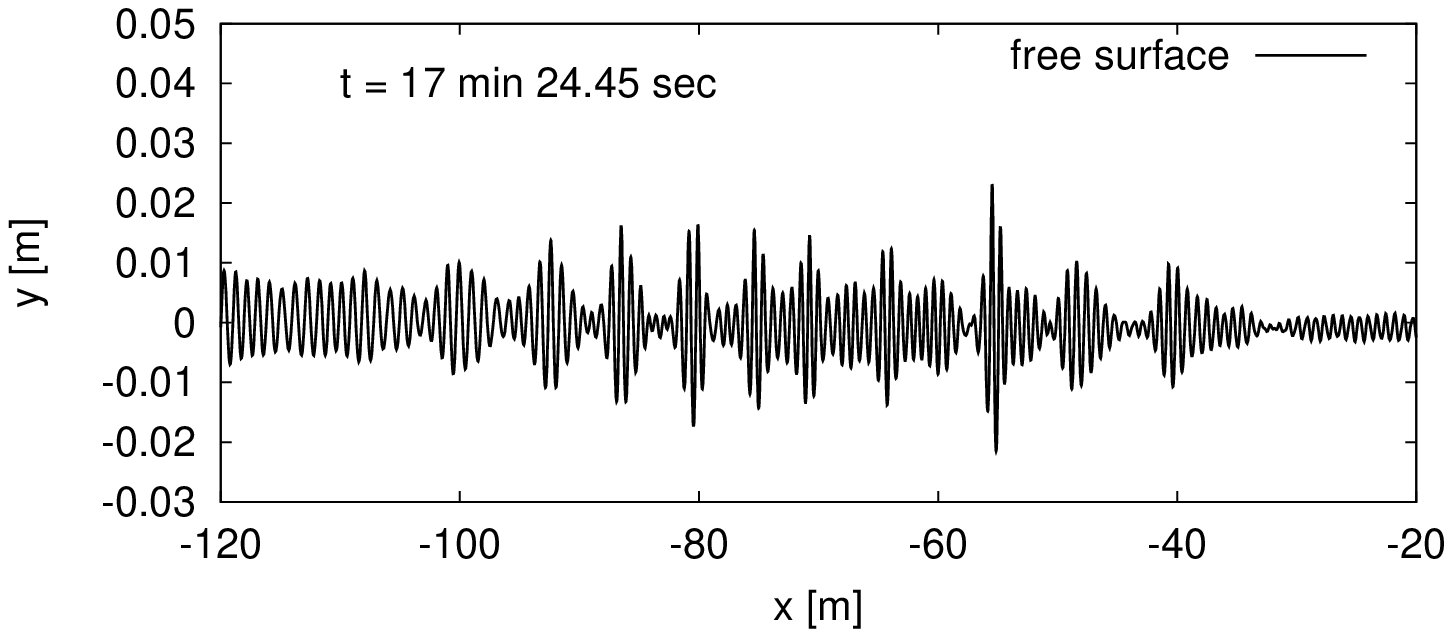,width=88mm}\\
   \epsfig{file=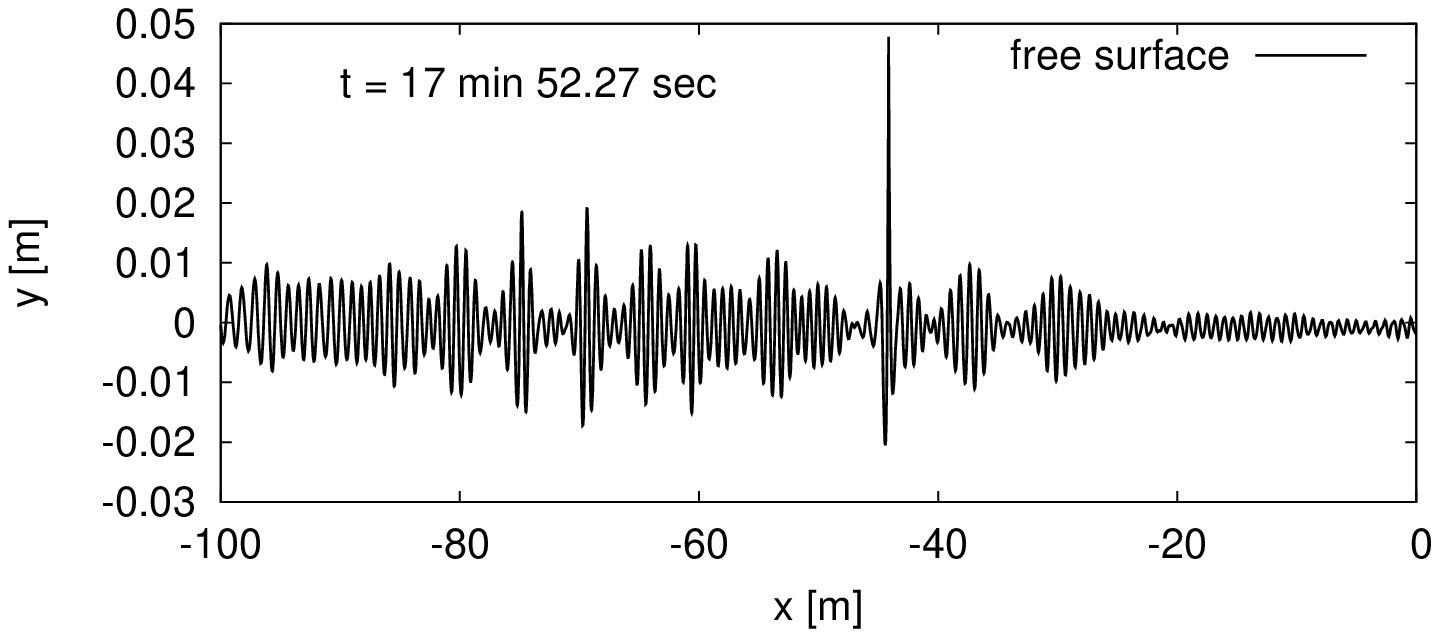,width=88mm}\\
\end{center}
\caption{Example of the appearance of a rogue wave in a
quasi-random wave field on the counter current.} 
\label{rogue_wave} 
\end{figure}

\begin{figure}
\begin{center}
   \epsfig{file=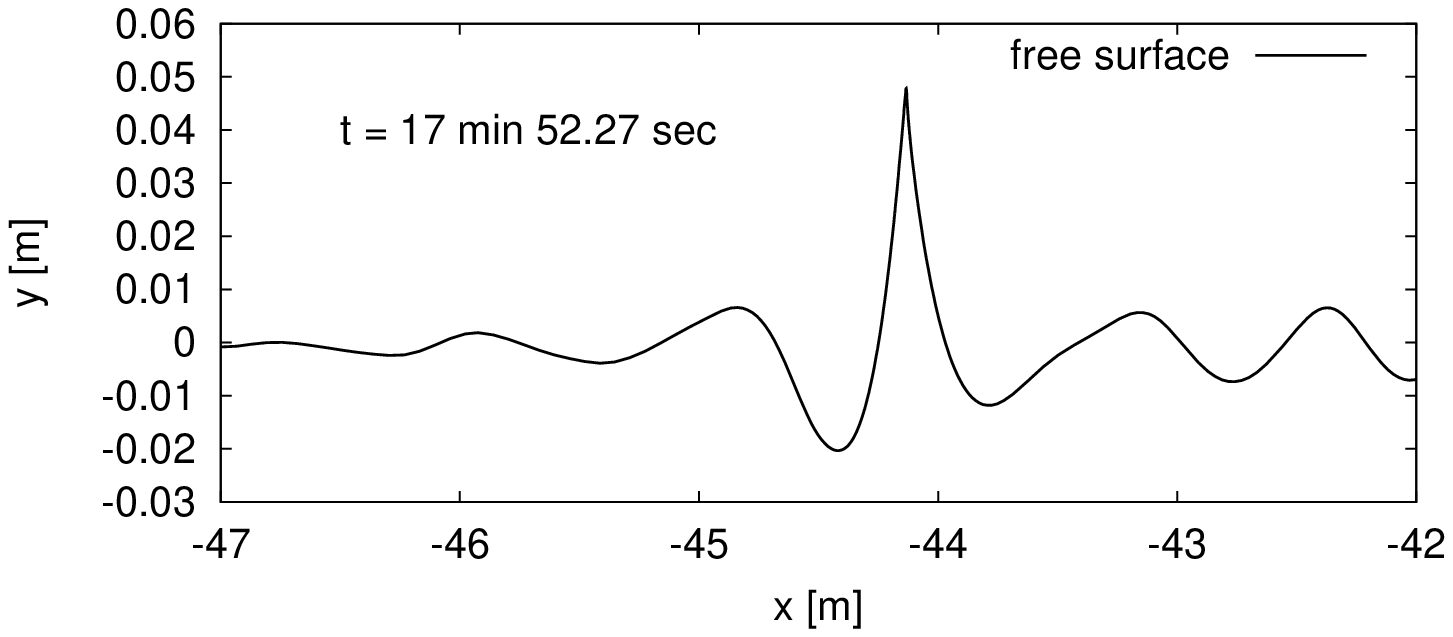,width=88mm}\\
\end{center}
\caption{Profile of the rogue wave from Fig. 4c.} 
\label{rogue_wave_profile} 
\end{figure}

{\bf Comparison with numerical experiment}.
It is easy to demonstrate that estimate (6) for the
amplitude of the formed anomalous waves exactly follows from Eq. (20) for the case of the propagation of a
weakly modulated wave from the region with the slow current $U_1$ to the region with the fast counter current
$U_2$. At the same time, this estimate can be tested by the direct numerical simulation of nonlinear waves on a
nonuniform current with the exact equations of the two-dimensional motion of the ideal fluid with free
surface, which are written in terms of so-called conformal variables. The results of such a simulation
described below show that Eq. (6) is quite reasonable (see Fig.1).

The (dimensionless) conformal variables parameterize the free boundary ($2\pi$ periodic in $x$) in terms of
the real functions $\alpha(t)>0$ and $\rho(\vartheta,t)$ as follows [29-31]:
\begin{equation}
X+iY=Z[\vartheta + i\alpha(t)+\{1+i\hat{\mathsf R}_\alpha\}\rho(\vartheta,t)],
\end{equation}
Here, $\hat{\mathsf R}_\alpha$ is the linear integral operator diagonal in the
discrete Fourier representation and ${\mathsf R}_\alpha(m)=i\tanh(\alpha m)$, where $m$ 
is the ordinal number of a Fourier harmonic. The given analytic function $Z(\zeta)$ with the
property $Z(\zeta+2\pi)=2\pi+Z(\zeta)$ determines the conformal mapping of a quite wide horizontal band in the
upper half-plane of the auxiliary complex variable $\zeta$
adjacent to the real axis $\mbox{Im }\zeta =0$ into the region in the
physical $(x,y)$ plane, and the real axis $\mbox{Im }\zeta =0$ parameterizes the bottom profile. 
In addition, the state of the system is characterized by the function $\psi(\vartheta,t)$ on
which the normal component of the velocity on the
surface depends. The functions $\rho(\vartheta,t)$ and $\psi(\vartheta,t)$ are
$2\pi$-periodic in the variable  $\vartheta$.

The exact compact expressions for the time derivatives $\rho_t(\vartheta,t)$, $\psi_t(\vartheta,t)$ and 
$\dot\alpha(t)$ have the form (see [32]; the difference here is in the presence of the parameter $s$, 
which is proportional to the average current velocity $U$)
\begin{eqnarray}
&&\!\!\rho_t=-\mbox{Re}[\xi_\vartheta(\hat {\mathsf T}_\alpha+i){\mathsf Q}],
\label{rho_t_alpha} 
\\
&&\!\!\psi_t=-\mbox{Re}[\Phi_\vartheta(\hat{\mathsf T}_\alpha\! +i){\mathsf Q}] 
-\frac{|\Phi_\vartheta|^2}{2|Z'(\xi)\xi_\vartheta|^2}
-g\,\mbox{Im\,}Z(\xi),
\label{psi_t_alpha}
\\
&&\!\!\dot\alpha(t)=-\frac{1}{2\pi}\int_0^{2\pi}{\mathsf Q}(\vartheta)d\vartheta,
\label{dot_alpha}
\end{eqnarray}
where
\begin{eqnarray}
&&\xi=\vartheta+i\alpha+(1+i\hat{\mathsf R}_\alpha)\rho, \\
&&\Phi_\vartheta=s+(1+i\hat{\mathsf R}_\alpha)\psi_\vartheta, \quad s=\mbox{const},\\
&&{\mathsf Q}=(\hat{\mathsf R}_\alpha \psi_\vartheta)/|Z'(\xi)\xi_\vartheta|^2.
\end{eqnarray}
The linear operator $\hat{\mathsf T}_\alpha$ is diagonal in the discrete Fourier representation: 
${\mathsf T}_\alpha(m)=-i\coth(\alpha m)$ for $m\not=0$ and ${\mathsf T}_\alpha(0)=0$.

The above equations were solved numerically using modern computer programs based on the fast Fourier
transform, as well as library functions of a complex variable in the C programming language. The function
$Z(\zeta)$ was taken in the form
\begin{equation}
\label{Z_zeta}
Z(\zeta)=\zeta -i\frac{6\pi}{400} + 
0.02\log\Big[\frac{1+0.96i\exp(i\zeta)}{1-0.96i\exp(i\zeta)}\Big].
\end{equation}
This function specifies the bottom profile shown in Fig. 2a (all lengths are rescaled to the spatial period
$L_x=400$ m). Before the beginning the calculations,  $\alpha_*$
and $\rho_*(\vartheta)$ corresponding to the steady state with the
average value $\langle\eta_0(x)\rangle\approx0$ were found numerically. In
this procedure, a certain current profile $U(x)$ was self-consistently established. Figure 2b shows the plot of
the function $U(x)$ at the parameter $s=-0.005$.

At the initial time, a sufficiently long modulated wave packet (which corresponded to initial data $\alpha(0)$
and $\rho(\vartheta,0)$ differing relatively little from $\alpha_*$ and  $\rho_*(\vartheta)$) 
was generated in the region with a slow current; then, it propagated against the current. When coming
to the strong counter current, modulation instability was developed and anomalous waves were formed in
agreement with the theory. Figure 3 exemplifies one such numerical experiment for $s=-0.005$, the initial
wavelength $\lambda_1=1.0$ m, steepness $\varepsilon_1=0.04$, and $N_1=8$.

Circles in Fig.1 show the maximum nonlinear
increase in the amplitude of the simulated wave at $\lambda_1=1.0$\,m, $\varepsilon_1=0.04$ and $N_1=8$,
 for $s=-0.006,-0.005,\dots,-0.002$. Squares in this figure are the data
at the parameters $\lambda_1=0.5$ m, $\varepsilon_1=0.032$ and  $N_1=10$
for $s=-0.004,-0.003,-0.002$. Agreement with the
theory is satisfactory, taking into account that the nonlinear Schr\"odinger equation is hardly applicable at the
final stage, when the energy is accumulated at the scale
of one or two wavelengths and the steepness somewhere exceeds 0.15-0.25. Moreover, the parametric
region near the instability threshold cannot be correctly studied in a numerical experiment, because very
long distances are necessary for reaching the maximum amplitude.

In conclusion, it is worth noting that, if a quasi-random sequence of groups of waves in the $I\sim 1$
regime, rather than a weakly modulated long wave
packet, comes to the counter current, appearing rogue
waves can be much higher than those predicted by
Eq. (6). The corresponding example is given in Figs. 4
and 5 (in this numerical experiment, $s=-0.004$,
$\lambda_1=1.0$ m, and the average steepness $\varepsilon_1 \approx  0.04$; 
the quasi-random wave field appears when the long packet
comes to the ``second round''). A reason for the appearance of higher anomalous waves is the attractive 
interaction of quasi-soliton coherent structures to which
typical wave groups are transformed when coming to
the fast counter current. The Akhmediev breather and
Eq. (6) based on it do not take into account the possible processes of fusion of quasi-solitons.

\end{document}